# Brief comment: Dicke Superradiance and Superfluorescence Find Application for Remote Sensing in Air

D. C. Dai*

*Physics Department, Durham University, South Road, Durham DH1 3LE, United Kingdom*
*Corresponding author: dechang.dai@durham.ac.uk*

This letter briefly introduces the concepts of Dicke superradiance (SR) and superfluorescence (SF), their difference to amplified spontaneous emission (ASE), and the hints for identifying them in experiment. As a typical example it analyzes the latest observations by Dogariu *et al.* (Science **331**, 442, 2011), and clarifies that it is SR. It also highlights the revealed potential significant application of SR and SF for remote sensing in air. © 2011 Optical Society of America
 *OCIS Codes: 020.1670, 270.1670, 260.7120, 010.0280, 280.1350*

In 1954 Dicke predicted the *cooperative* emission from a dense excited two-level system of gaseous atoms or molecules in population inversion in the absence of a laser cavity [1], soon after, Bonifacio and Lugiato termed it into two distinct forms [2]: Dicke Superradiance (SR) and Superfluorescence (SF). In a simplified picture, SR is from such an energy level prepared directly by laser into a quantum state which has an initial macroscopic dipole (coherence); whereas SF is from an initially incoherent energy level, which later on spontaneously builds-up a macroscopic dipole, in this case the system is started by normal fluorescent emission (FL), then gives rise to SF pulse, having a characteristic induction time ($\tau_D$) for the coherence development [2,3]. Both SR and SF are greatly limited by an opposed factor, the dephasing time ($T_2$ or $T_2^*$, defined by the inverse of linewidth of a corresponding optical transition containing homogeneous broadening only or inhomogeneous broadening respectively), which always acts to destroy the coherence [2,3].

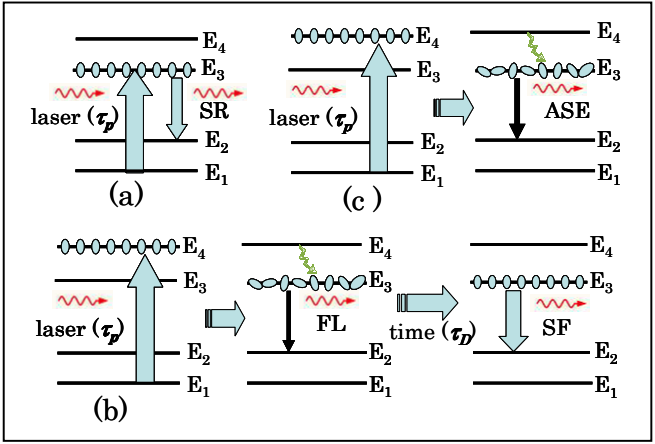

Fig. 1. (Color online) Illustration of SR (a), SF (b) and ASE (c) in a four-energy-level system in laser physics. Here, $E_3$ has relatively longer lifetime than $E_4$ and $E_2$, the fast relaxation from $E_4$ to $E_3$ leads to population accumulation on $E_3$ and results in a population inversion between $E_3$ and $E_2$. The property of emission from $E_3$ is determined by its dephasing time ($T_2^*$), for example, in (a) if $T_2^* \ll \tau_p$ it is not SR but ASE.

In contrast to SR and SF, amplified spontaneous emission (ASE) is the *collective* emission from such a similar but purely incoherent system. In ASE process, all the modes of spontaneous emission within gain can be amplified by stimulated emission in a single pass through the excitation volume, yet among them the mode with the maximum gain gradually wins others by competition effect, and finally results in coherent output at the end [3]. Fig. 1 illustrates three cases above.

In experiments, the phenomena of ASE, SR and SF are all appeared as stimulated emission or mirror-less lasing, their behaviors are quite similar to each other: they share the observable features based upon a clear threshold determined by population inversion, such as spectral line narrowing, exponentially growing intensity as pump power, and drastically shortened emission lifetime, directionality and polarization of emission beam *etc*, these usually bring much confusion to researcher in a test to identify a specific case of SR or SF from the more common cases of ASE [3].

Given the facts that SR and SF have been successfully observed predominantly in gaseous systems [4], since the first report in HF gas in 1973 [5], some hints are summarized here: in frequency domain, (i) doing energy level analyses as shown in Fig. 1, SR is from such an energy level resonantly populated by a pump laser, (ii) checking exponential index by fitting pump power dependent intensity growth, both SR and SF have an explicit index of 2.0 by definition in the case of single photon excitation, or 4.0 in two-photon excitation, for ASE it could be any number; then in time domain, (iii) looking for any quantum effect for SR and SF, such as the interference effect, quantum beats and ringing, which are the characteristics of the emission from coherent states, whereas ASE lacks of these; (iv) trying to resolve a time delay, $\tau_D$, between normal FL and lasing with a short pulse duration of pump laser, $\tau_p$. This $\tau_D$ is unique for SF, yet absent in both SR and ASE; (v) checking $T_2^*$ and comparing it to $\tau_p$, if $\tau_p$ is apparently longer than $T_2^*$, the emission is most possibly ASE rather than SR or SF; (vi) specifically, SF intensity exhibits quantum noise due to its triggering mechanism, and has other quantitative measures and criteria [6].

Very recently, Dogariu *et al.* report the backward lasing in air for remote sensing [7]. In their observations the 3P $^3$P level of Oxygen atom is resonantly populated by two-photon excitation with ultrashort laser pulse at 226 nm, the backward emission at 845 nm from this level has good directionality comparing to the lateral FL, the distinct power index of 4.2, which is consistent to the previous report [8], is in agreement with the theoretical value of 4.0 for two-photon excitation, these obviously fit the case (a) in Fig. 1; furthermore, by considering the pulse width is more than two orders of magnitude shorter than the FL lifetime, the evidence clearly points to SR, however, this has not been discussed or claimed by authors. By the instrumentally limited linewidth of 0.1 nm the estimated $T_2^*>24$ ps, this is on the same order of magnitude as $\tau_p=100$ ps used here, given the typical $T_2^*$ value of gaseous system at room temperature is on nanosecond range [4-6], which is apparently longer than $\tau_p=100$ ps, it can clearly clarify that the backward lasing at 845 nm in [7] is due to SR not ASE.

In an earlier report, Luo *et al.* observed the backward lasing at 357 nm from $N_2$ in an air plasma created by femtosecond laser, and simply interpreted it as ASE by the evidence of exponential growth and gain [9]. In this case it can be sure that the $N_2$ molecules are not resonantly excited by the pump laser at 810 nm used there, whereas the very short $\tau_p$ (42 fs) than typical $T_2^*$ value implies that the system should have sufficiently long time to develop coherence and radiates SF pulse, as the case (b) in Fig. 1. However, in order to confirm this to be SF, more detailed studies are required to find more characteristic features, like $\tau_D$, an explicit power index, quantum noise and possible quantum beating, ringing *etc*, together with quantitative analyses.

In summary, the backward Dicke SR and SF from laser plasma in atmospheric environment will potentially benefit the remote sensing techniques for detecting the molecules and gas pollutants, with their specific fingerprint emission lines, for example, 845 nm from Oxygen atom, or 337 nm from $N_2$ molecule. Also a fantastic idea arisen is that, these significant applications could be extended to study the atoms in the sun, a huge burning sphere.